\documentclass{article}

\usepackage{PRIMEarxiv}

\usepackage[utf8]{inputenc} 
\usepackage[T1]{fontenc}    
\usepackage{hyperref}       
\usepackage{url}            
\usepackage{booktabs}       
\usepackage{amsfonts}       
\usepackage{nicefrac}       
\usepackage{microtype}      
\usepackage{lipsum}
\usepackage{fancyhdr}       
\usepackage{graphicx}       
\graphicspath{{media/}}     

\usepackage{booktabs} 
\usepackage{array}
\usepackage{multirow}
\usepackage[export]{adjustbox}
\usepackage{pifont}
\usepackage{subfigure}
\usepackage{enumitem}

\usepackage[normalem]{ulem}
\usepackage{setspace}

\title{Feasibility of Interactive 3D Map for Remote Sighted Assistance
}

\author{
  Jingyi Xie\textsuperscript{1}, 
  Rui Yu\textsuperscript{1,}\thanks{First two authors contributed equally to this work.}\hspace{1mm},
  Sooyeon Lee\textsuperscript{2}, 
  Yao Lyu\textsuperscript{1}, 
  Syed Masum Billah\textsuperscript{1}, 
  John M. Carroll\textsuperscript{1} 
  \\
  {1} College of Information Sciences and Technology, Pennsylvania State University
  \\
  {2} School of Information, Rochester Institute of Technology
}

\begin{document}
\maketitle

\begin{abstract}
Remote sighted assistance (RSA) has emerged as a conversational assistive technology, where remote sighted workers, i.e., agents, provide real-time assistance to users with vision impairments via video-chat-like communication. Researchers found that agents' lack of environmental knowledge, the difficulty of orienting users in their surroundings, and the inability to estimate distances from users' camera feeds are key challenges to sighted agents. To address these challenges, researchers have suggested assisting agents with computer vision technologies, especially 3D reconstruction. This paper presents a high-fidelity prototype of such an RSA, where agents use interactive 3D maps with localization capability. We conducted a walkthrough study with thirteen agents and one user with simulated vision impairment using this prototype. The study revealed that, compared to baseline RSA, the agents were significantly faster in providing navigational assistance to users, and their mental workload was significantly reduced -- all indicate the feasibility and prospect of 3D maps in RSA.
\end{abstract}

\keywords{People with visual impairments \and Remote sighted assistance \and Computer vision \and 3D map \and Augmented reality \and Navigation}

\section{Introduction}
\begin{figure}[!tb]
\centering
\includegraphics[width=0.98\linewidth]{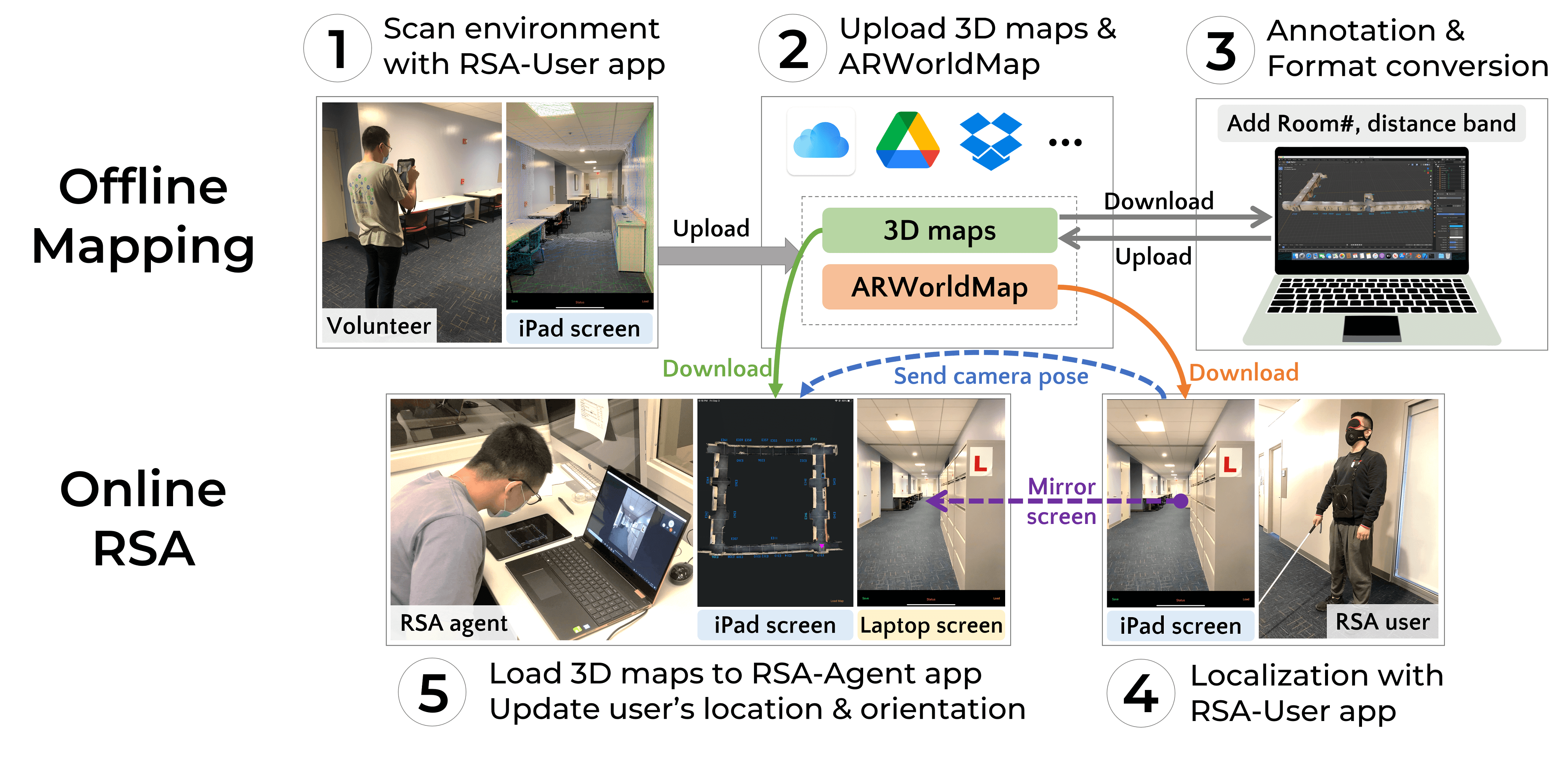}
\vspace{-0.1cm}
\caption{Overview of our RSA prototype with interactive 3D maps.}
\label{fig:workflow}
\end{figure}

Navigating indoor or outdoor spaces independently remains a significant challenge for people with visual impairments (PVI)~\cite{survey_of_navigation_aid_2019}, as they need to interact with a world they cannot see.
As such, they rely on various means, including their other sensory abilities~\cite{cognitive_colalge_1993, brain_plasticity}, orientation and mobility (O\&M) skills~\cite{wiener2010foundations}, mobility aids (e.g., traditional white cane and guide dog), and assistive technologies. 
Among many assistive technologies, smartphone sensor-based navigation apps~\cite{jain2014path, sato2017navcog3, BlindSquare2020, Soundscape2020, SeeingEye_GPSTM2020, Autour2020, Asakawa2018indoor} are the most widely deployed. 
Recently, computer vision-based navigation assistance~\cite{li2016isana, kayukawa2019bbeep, guerreiro2019cabot} and more recently, augmented reality-based navigation apps~\cite{yoon2019leveraging, ARKit-based-navigation} are also becoming mainstream.

Despite being high-tech, many of these assistive technologies are either tested in constrained laboratory scenarios, or are expensive, unreliable at times, cumbersome, or hard to distribute on a mass scale~\cite{wiener2010foundations, fallah2013indoor}.
Therefore, having a sighted in-person assistance is still arguably the most efficient method for PVI. 
Unfortunately, such a reliance on sighted persons, who are usually the family members or friends of a visually impaired individual, is burdensome to them, and sometimes they are not available to assist~\cite{gurari2018vizwiz}.
For these reasons, PVI start relying on \emph{remote} sighted assistants, who are always available and may not have any prior acquaintance with them. 
This form of assistance is generally called \emph{remote sighted assistance (RSA)} service~\cite{lee2020emerging}, in which a visually impaired user (namely, \emph{RSA user} or simply \emph{user}) first establishes a video connection with a remote sighted assistant (namely, \emph{RSA agent} or simply \emph{agent}), who then interprets the video feed from the user's smartphone camera and converses with the user to provide assistance as needed or requested.
Recently, a number of RSA services came out of academia, e.g., VizWiz~\cite{bigham2010vizwiz}, BeSpecular~\cite{holton2016BeSpecular}, Crowdviz~\cite{holton2015crowdviz}, as well as of industry, e.g., TapTapSee~\cite{TapTapSee2020}, BeMyEyes~\cite{BeMyEyes2020}, Aira~\cite{Aira2020}.

As RSA services have gained popularity, a growing number of researchers started to study different aspects of these services. Kamikubo et al.~\cite{kamikubo2020support} and Lee et al.~\cite{lee2018conversations, lee2020emerging, Caroll2020Human, iui} studied the interaction paradigm between RSA agents and visually impaired users and reported several challenges for RSA agents.
Some of the key challenges include agents' lack of confidence due to unfamiliarity of the users' current physical environment, not having indoor maps with finer details, continuously tracking and orienting the user in the map and within his or her surroundings, estimating objects' depth in users' camera feed and describing those in realtime, 
and developing mutual trust and synergy with the users.

Fortunately, these challenges are well studied in computer vision (CV) and AI literature, under the umbrella of indoor map construction and localization~\cite{CadenaCCLSN0L16}, depth estimation~\cite{SaxenaSN09, EigenPF14}, object tracking~\cite{SmeuldersCCCDS14}, visual navigation~\cite{Bonin-Font08, ZhuMKLGFF17}, 
and scene understanding~\cite{OlivaT01, RenHG017}.
However, these CV problems are still not perfectly solved by state-of-the-art algorithms in applications such as autonomous vehicles. 
We believe the challenging CV problems could be simplified if we shift the focus from assisting visually impaired users (as done in SeeingAI~\cite{SeeingAI2020}, BlindSquare~\cite{BlindSquare2020}) to assist remote sighted assistants. 
This is because unlike the former, the latter user group could override any potential mistakes made by CV.
Additionally, we believe the challenges in current RSA services could be lowered if RSA agents could delegate some of their workloads to a CV-mediated system.
Thus, CV-supported RSA can serve as a promising navigational aid for PVI.

We evaluated the desirability of CV-mediated RSA in our prior work~\cite{c4vtochi}. We developed a low-fidelity prototype showcasing potential applications for CV to support RSA interactions and reviewed it with professional and trained RSA agents. RSA professionals were engaged by, and reacted constructively to the prototyped CV concepts and design ideas, including 3D maps, real-time localization, annotations, and distance bands. We found that the proposed design ideas have the potential to improve the RSA practice by reducing the agent's cognitive load and enhancing the agent's ability to stay ahead. Although we received positive feedback from RSA professionals on the low-fidelity prototype, they were unable to actually interact with the prototyped CV support.

In this paper, we propose to design a higher-fidelity prototype to study the feasibility of the CV-mediated RSA system based on interactive 3D maps. In line with identified design objectives, we implemented an RSA prototype including a pair of mobile apps, namely RSA-User and RSA-Agent. Figure~\ref{fig:workflow} shows the workflow of our prototype which comprises two phases: offline mapping (Step 1-3) and online RSA (Step 4-5). In the offline mapping phase, we create 3D maps with the RSA-User app and add annotations. In the online RSA phase, agents can access an interactive 3D map of the user's surroundings in addition to the live camera feed. The user's position and orientation are updated on the 3D map in real-time.
The CV-supported RSA mobile apps are implemented with Apple's ARKit~\cite{arkit} and can achieve accurate real-time localization on 3D maps with the iOS devices equipped with a LiDAR scanner\footnote{Up to August 2021, the iPhone 12 Pro, Pro Max, 2020 and 2021 iPad Pro featured a LiDAR scanner.}. 

We conducted a walkthrough study to evaluate our RSA prototype by recruiting 13 sighted participants as RSA agents. Considering PVI are vulnerable in the COVID-19 pandemic~\cite{armitage2020covid}, this feasibility research did not take in any PVI for safety considerations. One sighted volunteer wearing a blindfold acted as the user with simulated visual impairment throughout the study.
In the experiments, each agent-user pair was assigned four navigational tasks in two different testing areas inside a campus building. Two tasks were completed with interactive 3D maps, while the other two were performed with static 2D maps for comparison. We also designed two types of tasks to examine the effects of annotations and fine-grained details in 3D maps.
We recorded the task completion time and NASA-TLX scores of each agent. The quantitative analysis showed that 3D maps could significantly reduce the completion time and mental workload compared with 2D maps. 
The results of this walkthrough study are encouraging in terms of enhancing RSA services with interactive 3D maps. Our prototype also provides a concrete implementation to demonstrate the feasibility of interactive 3D maps for RSA.

\section{Background and Related Work}

\subsection{Navigational Techniques for People with Visual Impairments}
Over the last 70 years, researchers proposed many prototypes to aid PVI in outdoor and indoor navigation.
In this section, we only review a subset of such prototypes that are widely used and run on smartphones (for a chronological review, see Real and Araujo~\cite{survey_of_navigation_aid_2019}).
Outdoor navigation apps, such as NavCog~\cite{NavCog2016}, BlindSquare~\cite{BlindSquare2020}, SeeingEyeGPS~\cite{SeeingEye_GPSTM2020}, Soundscape~\cite{Soundscape2020}, and Autour~\cite{Autour2020}, rely on GPS sensors for localization and commercial map services (e.g., Google Map, OpenStreet Map) for wayfinding.
PVI can navigate large distances using these apps, but struggle to find the last-few-meters~\cite{saha2019closing} due to a wide margin of error in GPS accuracy ($\pm$5m~\cite{GPS2020}).

For indoor navigation, these apps are not reliable because of the weaker GPS signal strength indoors and not having sufficiently detailed indoor map data~\cite{rodrigo2009robust, li2010indoor}.
To overcome these limitations, researchers have proposed many techniques that support indoor localization by fusing available smartphone sensors (e.g., Bluetooth~\cite{sato2017navcog3}, Infrared~\cite{legge2013indoor}, NFC~\cite{ganz2014percept}, RFID~\cite{ganz2011percept}, sonar~\cite{DokmanicAcoustic2013}, and camera)
or construct indoor maps by understanding the semantic features of the environment (for a complete list, see Elmannai and Elleithy~\cite{Elmannai2017SensorBasedAD}).
Unfortunately, these solutions require additional deployment and maintenance effort, as well as cost for setting up databases of floorplan~\cite{fallah2012user} and structural landmarks~\cite{bai2014landmark, perez2017assessment}. Some solutions also require users to carry specialized devices (e.g., an IR tag reader~\cite{legge2013indoor}).
Other solutions, such as LuzDeploy~\cite{gleason2018crowdsourcing}, propose to engage non-expert volunteers for maintaining the deployed infrastructure. 
For these reasons, no single indoor navigation system is widely deployed. 
In this work, we envision that combining navigation techniques and human-assisted RSA service have the potential to address indoor navigation challenges in a collaborative manner.


\subsection{Remote Sighted Assistance Services for People with Visual Impairments}

The implementation of various RSA services differs in three key areas: (i) the communication medium between users and remote sighted assistants. Earlier prototypes used audio~\cite{petrie1997mobic}, images~\cite{kutiyanawala2011teleassistance, bigham2010vizwiz}, one-way video using portable digital cameras~\cite{garaj2003system, bujacz2008remote}, or webcams~\cite{bujacz2008remote}, whereas the recent ones are using two-way video with smartphones~\cite{baranski2015field, holmes2015iphone, BeMyEyes2020, Aira2020}; 
(ii) the instruction form, e.g., via texts~\cite{lasecki2013chorus}, synthetic speech~\cite{petrie1997mobic}, natural conversation~\cite{baranski2015field, BeMyEyes2020, Aira2020}, or vibrotactile feedback~\cite{scheggi2014remote, chaudary2017tele};
and (iii) localization technique, e.g., via GPS-sensor, crowdsourcing images or videos~\cite{rafian2017remote, bigham2010vizwiz, lasecki2011real, zhong2015regionspeak}, fusing sensors~\cite{rafian2017remote}, or using CV as discussed in the next subsection.

Researchers examined the feasibility of crowdsourced RSA services (e.g., TapTapSee~\cite{TapTapSee2020}, BeMyEyes \cite{BeMyEyes2020}), and concluded that this is a promising direction to tackle navigation challenges for blind users~\cite{avila2016remote, brady2015crowdsourcing}. 
However, they commented on the issue of crowdworkers not being available at times~\cite{burton2012crowdsourcing}.
Nguyen et al.~\cite{nguyen2018improvement} and Lee at al.~\cite{lee2020emerging} studied a paid RSA service, Aira~\cite{Aira2020}. They reported that unlike crowdworkers, Aira agents are always available and trained in communication terminology and etiquette. 
In this paper, we propose a new RSA prototype to study the feasibility of interactive 3D maps with CV techniques to improve existing RSA services for indoor navigation.

\subsection{Use of Computer Vision in Navigation for People with Visual Impairments}
\label{cv-rsa}
Budrionis et al.~\cite{budrionis2020smartphone} reported that CV-based navigation apps on smartphones are a cost-effective solution for indoor navigation. 
A major focus on CV-based approach is how to make visual information more accessible through recognizing objects~\cite{Zientara2017thirdEye}, obstacles~\cite{presti2019watchout}, color-codes or landmark (e.g., storefronts~\cite{saha2019closing}), or through processing of tags such as barcodes~\cite{Tekin2010Barcode}, QR codes~\cite{elgendy2019indoor}, and RFID~\cite{McDaniel2008RFID}.
Extending this focus, researchers have proposed indoor positioning and navigation systems~\cite{ko2017vision, legge2013indoor, manduchi2010blind}.
However, Saha et al.~\cite{saha2019closing} concluded that for a deployable level of accuracy, using CV techniques alone is not sufficient yet.
In this work, we use CV and 3D maps to assist sighted assistants rather than PVI, who could be vulnerable to inaccuracies of CV systems.

Another line of work is to develop autonomous location-aware pedestrian
navigation systems. These systems combine CV with specialized hardware (e.g., suitcase~\cite{kayukawa2019bbeep}, wearable CV device~\cite{liu2018augmented}) and sensors (e.g., Lidar~\cite{guerreiro2019cabot}, Bluetooth~\cite{guerreiro2019airport}), and support collision avoidance.
However, their real-world adaptability is still questionable, as Banovic et al.~\cite{banovic2013uncovering} commented that navigation environments in real-world are dynamic and ever-changing. 

Lately, researchers are exploring the potential of augmented reality (AR) toolkit in indoor navigation. This toolkit is built into smartphones (e.g., ARKit~\cite{arkit} in iOS devices, ARCore~\cite{arcore} in Android devices), thus has the possibility of a widespread deployment~\cite{ARKit-based-navigation}. 
Clew demonstrated the potential of using AR toolkit to localize blind users on a pre-recorded route with acceptable accuracy~\cite{yoon2019leveraging}.
Verma et al.~\cite{verma2020indoor} reported that an AR-based navigation system could provide a better user experience than traditional 2D maps. 
Brata et al.~\cite{brata2020comparative} found that sighted participants rate 2D digital map over a location-absed AR interface for efficiency and dependability, but AR interface over 2D digital map for hedonic qualities and overall performance. 
Fusco et al.~\cite{fusco2020indoor} also reported that with ARKit, PVI do not need to keep the camera towards a target to recognize it. 
Troncoso Aldas et al.~\cite{troncoso2020aiguide} proposed an ARKit-based app, AIGuide, to help PVI to recognize and localize objects. 
Unlike prior work, in this paper, we explore the viability of utilizing AR toolkit and 3D maps, only to assist sighted agents instead of PVI.

\subsection{Collaboration between Human and AI}
Despite the recent advancements in CV and automatic scene understanding, 3D reconstruction from video stream remains a challenge. The performance of existing systems is impacted by various factors, such as motion blur, change of light, scale, and orientation~\cite{JafriAAF14}. As a solution to this limitation, interactive, hybrid approaches that involve human-AI collaboration have been studied. Brady et al.~\cite{BradyMZWB13} offered users several options for answer sources, including crowd-sourced workers, to manually identify objects unrecognized by CV algorithms. An interesting variation of hybrid approaches is the human-in-the-loop framework. Branson et al.~\cite{BransonWSBWPB10} used human responses to questions posed by the computer to drive up recognition accuracy while minimizing human effort. Some researchers studied interactive 3D modeling in which humans provide guidance by drawing simple outlines or scribbles~\cite{SinhaSSAP08, KowdleCGC11}. 
Our approach is in line with the above research, but we explore the human-AI collaboration design space leveraging CV and 3D maps to support sighted human assistants.

\section{Prototype Design}

\subsection{Design Objectives}

In this study, we explore the feasibility of interactive 3D maps to enhance RSA. Towards this end, we propose to design a new RSA prototype for assessment. According to prior work~\cite{c4vtochi}, we identified several objectives for designing the new RSA service:
\begin{itemize}
  \item \textbf{Basic RSA function.} Existing RSA services (e.g., Aira, BeMyEyes) mainly support live video chat between the user and agent. 
  Aira agents emphasize the importance of the live video feed and consider it as the \textit{``lifeline of information''}.
  Although web-based information (e.g., maps and satellite images) is useful, live video and audio feeds from PVI's mobile devices are the only resource of information that is real-time and guaranteed to be accurate and up-to-date~\cite{c4vtochi}. 
  The basic video chat function is common and can be executed with off-the-shelf tools. For example, Kamikubo et al.~\cite{kamikubo2020support} realized the user-agent interaction with a video conferencing system.

  \item \textbf{Interactive 3D maps.} To address the lack of maps in indoor navigation, the RSA agents expect to learn the environmental context of the users from 3D maps in an interactive manner similar to on a web mapping platform (e.g., Google Maps) they have employed in existing RSA services for outdoor navigation~\cite{c4vtochi}. The 3D maps also need to allow agents to change the scale and view via common interactions (e.g., zooming, panning, and rotating) to find both general and detailed information.
  In practice, 3D maps are usually saved as polygon meshes or point clouds in various formats (e.g., \texttt{.obj}, \texttt{.dae}, \texttt{.ply}), preferably with colored texture.
  
  \item \textbf{Real-time localization.} Real-time localization has been widely applied in wayfinding apps for people with and without visual impairments. These apps (e.g., Google Maps~\cite{google_map_app}) rely on GPS for localization, which is a mature technology for outdoor navigation. However, the weaker GPS signal strength and low accuracy render these apps unreliable in the indoor environment. Informed by prior study~\cite{c4vtochi}, continuously updating the user's position and orientation on a 3D map is desired by professional RSA agents for indoor navigation. To simplify the system, we prefer to realize real-time localization only using standalone mobile devices (phones or tablets) without extra environmental infrastructure (e.g., Bluetooth beacon~\cite{NavCog2016}, RFID tags~\cite{ganz2014percept}). 3D maps using augmented reality (AR) technology have shown the potential to address this problem with smartphone~\cite{yoon2019leveraging}. With the rapid development of AR toolkits (e.g., Apple's ARKit, Google's ARCore) and more powerful sensors (e.g., LiDAR scanner) equipped in mobile devices, it is feasible to achieve reliable real-time localization with AR technology and mobile devices.

  \item \textbf{Annotation.}
  Annotations on 3D maps, such as room numbers in office buildings and aisles and sections in grocery stores, provide agents with more spatial details and inform them of possible paths. Agents considered annotations \textit{``very helpful just to get a sense for the location and the store''} and help them to guide PVI more efficiently~\cite{c4vtochi}. 
  Moreover, supposing the destination is a specific room, the annotation of the room number on 3D map will greatly ease this challenging indoor navigation task.
  
  \item \textbf{Distance Bands.}
  It is difficult for agents to estimate distance through a live video feed, especially considering differences in camera height and angle. Distance band is one of the methods to present distance information as a grid overlaid on maps. Agents gave positive feedback about this feature because it enhances their ability to stay ahead of PVI and prepare for approaching obstacles~\cite{c4vtochi}. Compared with other distance measurements overlaid on the video feed, distance bands have the advantage of not obscuring the \textit{``lifeline''} real-time video feed. In this prototype, we draw distance bands on the 3D maps with an interval of 10 feet. 
  
  \item \textbf{Dashboard with extra information.}
  Professional agents indicated that they use split-screen or multiple monitors and \textit{``sometime will have several maps open''} for reference to identify points of interest~\cite{c4vtochi}. To support their habit, the new RSA service will follow the split-screen dashboard design. Specifically, the 3D map as an additional information source will be displayed on a separate screen from the live video feed. The manipulation will mostly be performed on the interactive 3D map screen.
  
\end{itemize}

\subsection{Prototype Workflow}

In accordance with the design objectives, we prototyped a new RSA service with a pair of iOS apps, namely RSA-User and RSA-Agent. Figure~\ref{fig:workflow} shows the workflow of the proposed RSA prototype. The prototype consists of two phases: offline mapping and online RSA. In the offline mapping phase, we create 3D maps and save them in cloud storage. In the online RSA phase, the agent assists the user with both video chat and 3D maps. We briefly introduce the five steps in Fig.~\ref{fig:workflow} as follows and will detail the implementation in Section~\ref{sec:implementation}.

\begin{itemize}
  \item \textbf{Offline mapping}
    \begin{enumerate}
      \item A sighted volunteer scans the indoor environments of interest by an iPad Pro with the RSA-User app.
      \item The sighted volunteer saves the scanned 3D maps and ARWorldMap (containing mapping state and anchors for relocalization supported by Apple's ARKit~\cite{arkit}) files to cloud storage (e.g., Google Drive).
      \item The sighted volunteer downloads the 3D map files to a laptop, adds annotations (e.g., room numbers and distance band) to the maps, converts the map files to SceneKit-readable format, and uploads to cloud storage.
    \end{enumerate}
  \item \textbf{Online RSA}
    \begin{enumerate}
      \setcounter{enumi}{3}
      \item The user downloads the ARWorldMap file of the current surroundings to the RSA-User app for relocalization. The RSA-User app will continuously send the camera pose (position and orientation) to the RSA-Agent app.
      \item The agent loads the 3D map of the user's current area and views the user's real-time location in the RSA-Agent app. Meanwhile, the user's iPad screen is mirrored to the agent's laptop as the live camera feed.
    \end{enumerate}
\end{itemize}

\section{Implementation}
\label{sec:implementation}

\begin{figure}[!tb]
\setlength{\abovecaptionskip}{-0.03cm}
\centering{
\subfigure[3D map]
{
   \label{fig:3d_map-1A}
   \includegraphics[width=0.289\linewidth]{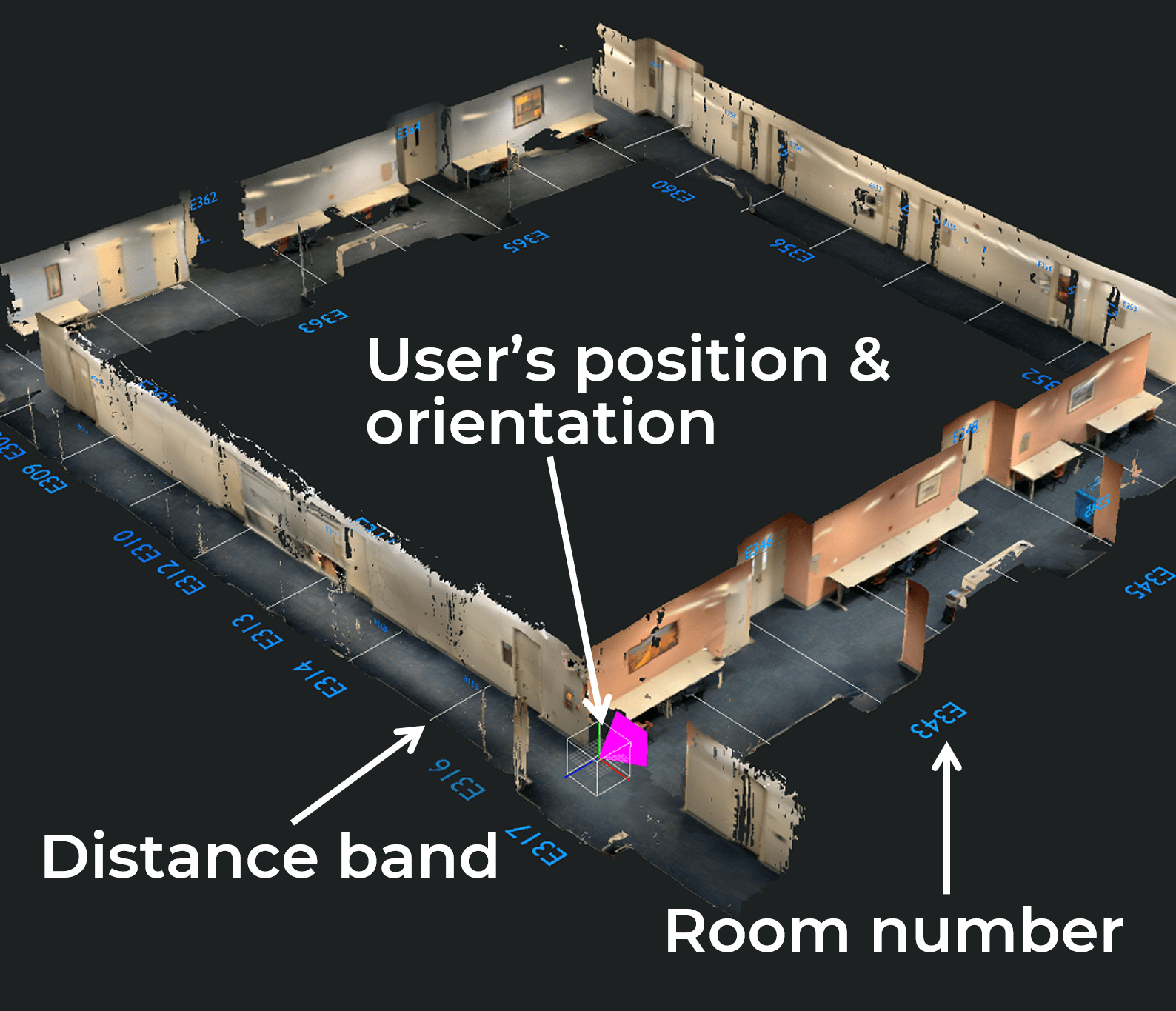}
}\hspace{-0.1cm}
\subfigure[Top-down view]
{
   \label{fig:3d_map-1B}
   \includegraphics[width=0.289\linewidth]{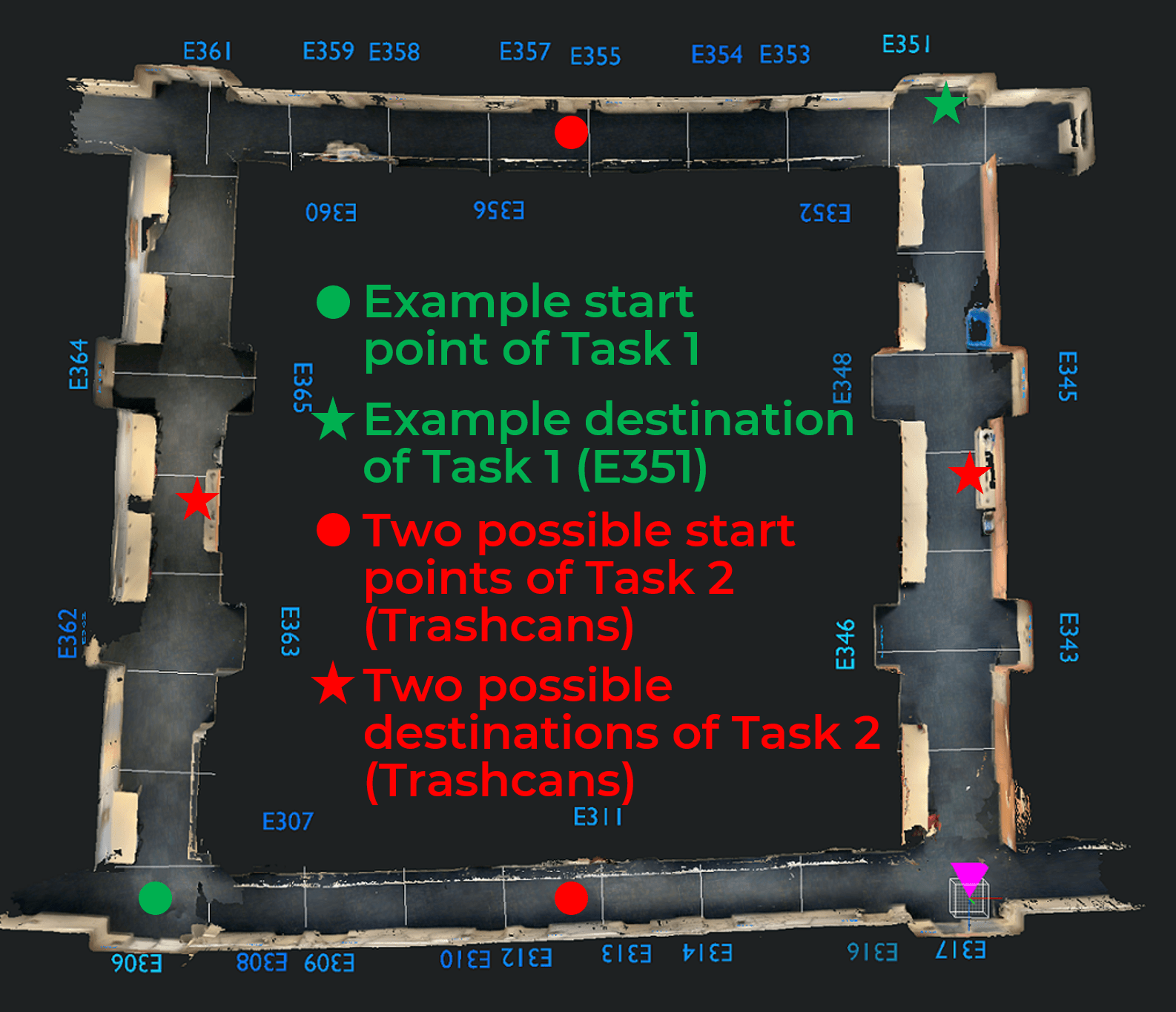}
}\hspace{-0.1cm}
\subfigure[Trashcans]
{
   \label{fig:trashcan}
   \includegraphics[width=0.139\linewidth]{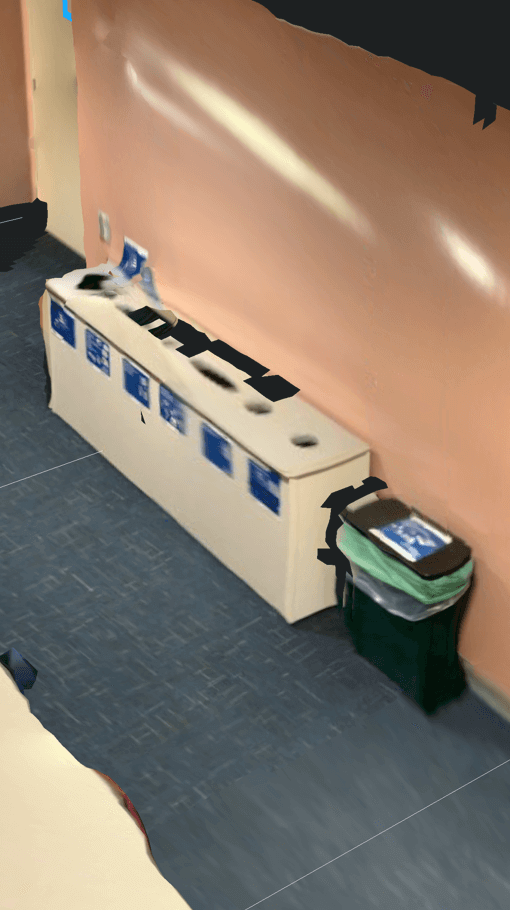}
}\hspace{-0.1cm}
\subfigure[2D map]
{
   \label{fig:2d_map-1}
   \includegraphics[width=0.242\linewidth]{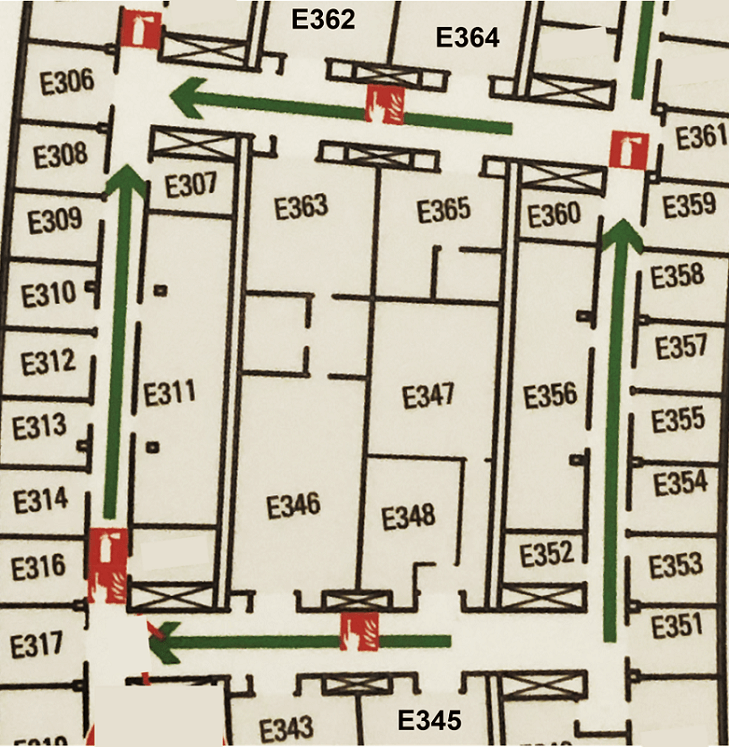}
}
}
\caption{Maps of the first testing area: (a) 3D map with annotations and real-time localization; (b) Top-down view of 3D map and the start points and destinations of Task 1 and Task 2; (c) Trashcans in 3D map (zoom in); (d) 2D map.}
\label{fig:maps-1}
\end{figure}
\setlength{\floatsep}{0.5cm}

\begin{figure}[!tb]
\setlength{\abovecaptionskip}{-0.03cm}
\centering{
\subfigure[3D map]
{
   \label{fig:3d_map-2A}
   \includegraphics[width=0.289\linewidth]{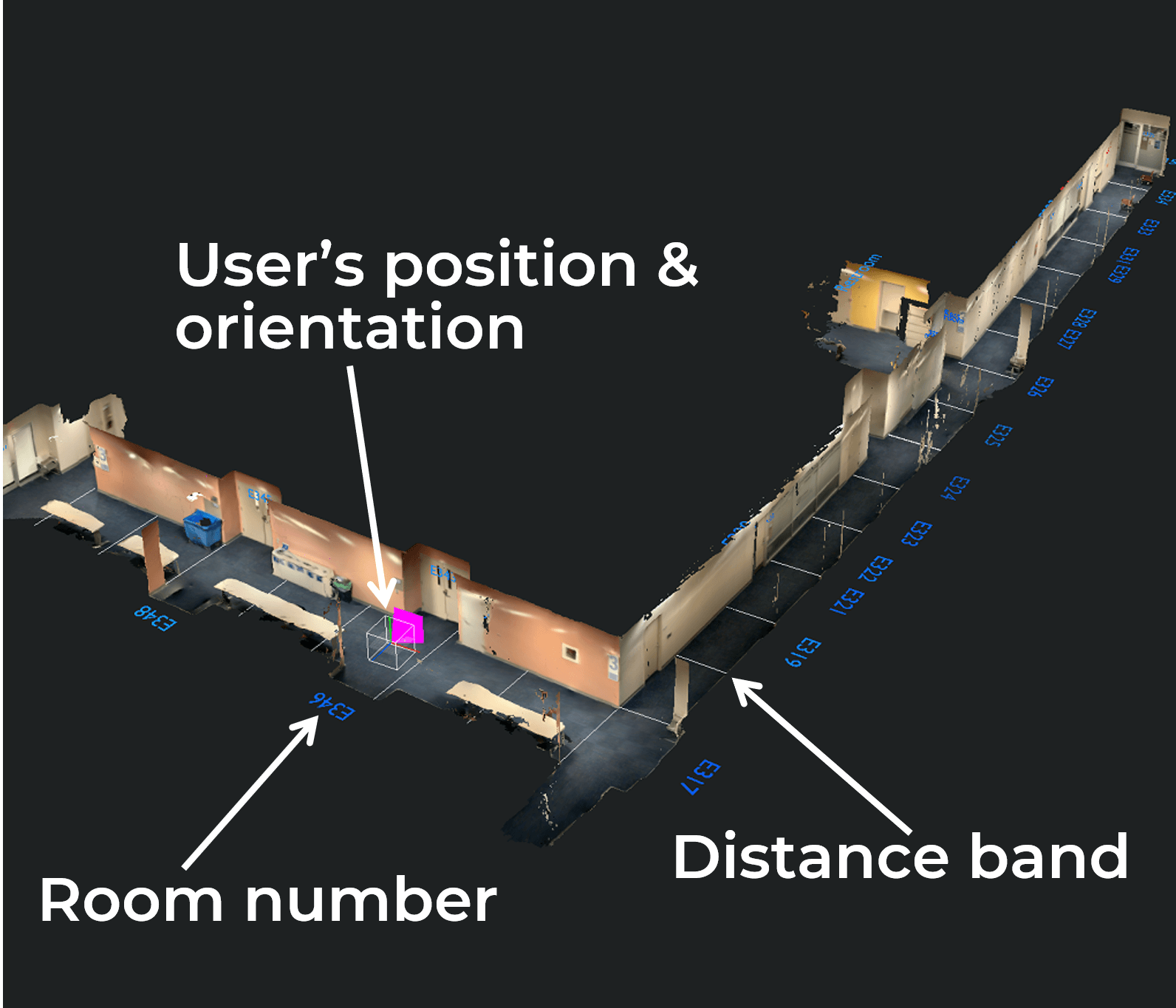}
}\hspace{-0.1cm}
\subfigure[Top-down view]
{
   \label{fig:3d_map-2B}
   \includegraphics[width=0.289\linewidth]{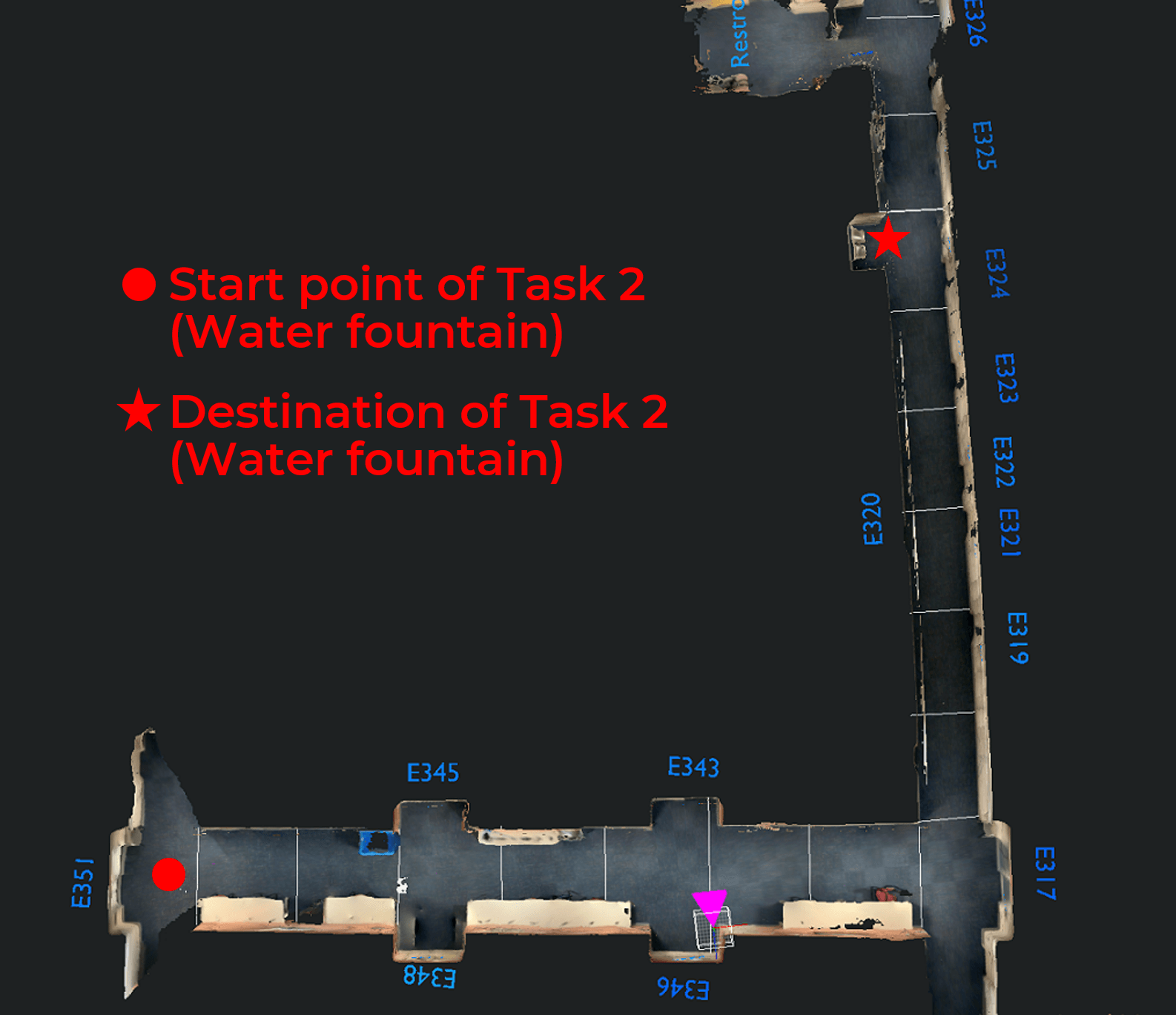}
}\hspace{-0.1cm}
\subfigure[Water fountain]
{
   \label{fig:water_fountain}
   \includegraphics[width=0.139\linewidth]{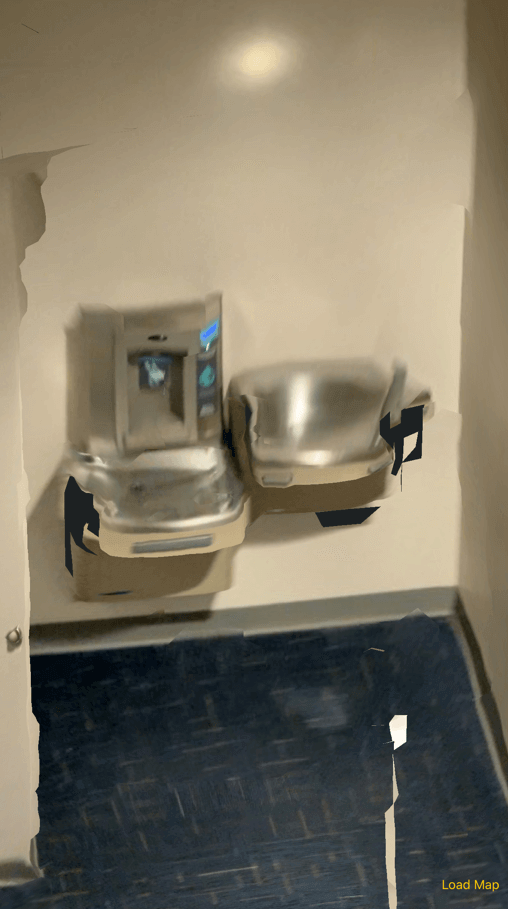}
}\hspace{-0.1cm}
\subfigure[2D map]
{
   \label{fig:2d_map-2}
   \includegraphics[width=0.242\linewidth]{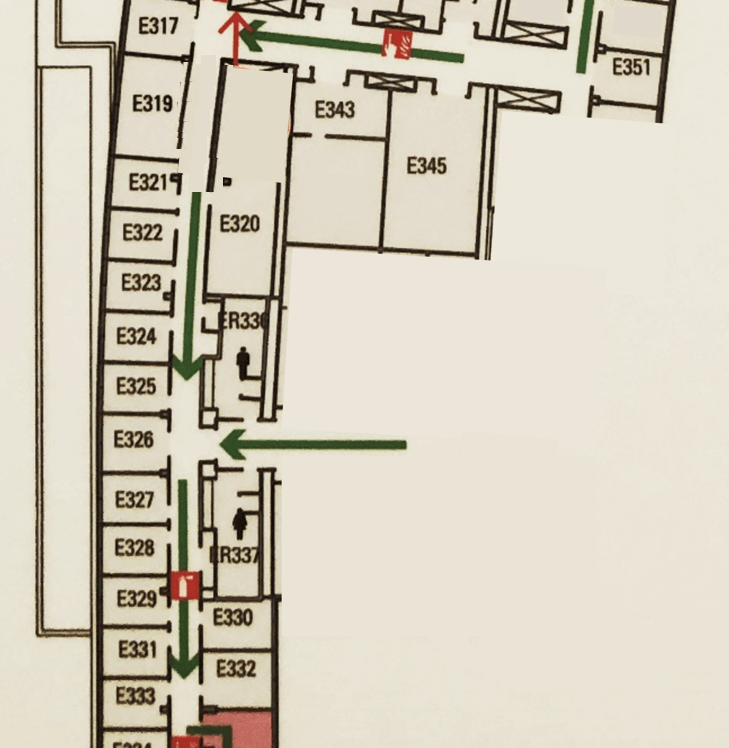}
}
}
\caption{Maps of the second testing area: (a) 3D map with annotations and real-time localization; (b) Top-down view of 3D map and the start points and destinations of Task 2; (c) Water fountain in 3D map (zoom in); (d) 2D map.}
\label{fig:maps-2}
\end{figure}

We implemented RSA-User and RSA-Agent apps with Apple's ARKit~\cite{arkit}, RealityKit~\cite{realitykit}, and SceneKit~\cite{scenekit}. All the codes were developed in Swift and Xcode version 12.4. The RSA-User app only supports Apple mobile devices equipped with a LiDAR scanner.

\subsection{Offline Mapping}
\subsubsection{Build 3D Maps}
There are two main reasons why we adopt Apple iOS devices for 3D mapping. First, Apple's ARKit facilitates camera relocalization by encapsulating the mapping state and anchors into an ARWorldMap object. This feature is crucial to the accurate real-time localization of RSA users in 3D maps.
Second, the new Apple mobile devices featuring a LiDAR scanner, along with ARKit, support powerful 3D scene understanding capabilities and reconstructing high-quality polygon mesh as 3D maps.

In our experiment, the sighted volunteer used the RSA-User app to scan areas of interest with a 2020 iPad Pro. During scanning, the RSA-User app continuously constructed a 3D mesh and displayed it overlaying on surfaces in real scenes. Meanwhile, ARKit constantly updated an ARWorldMap object, which contains the space-mapping state and ARAnchor objects for relocalization the next time.

When completing the scanning, the sighted volunteer tapped the ``Save'' button and chose the cloud storage to save both the 3D mesh and ARWorldMap. The 3D mesh was extracted from the ARMeshAnchor object of the current frame. The coordinates of the mesh vertices were converted to the same coordinate system as ARWorldMap with an anchor-to-world transform matrix. Then, the 3D mesh was exported to \texttt{.obj} (Wavefront object) file with Model I/O~\cite{modelio}. In RSA-User app, we defined a new file extension \texttt{.arworldmap} to save ARWorldMap object.
ARMeshAnchor only contains an untextured mesh. In the preliminary test, we found the 3D maps without textures could not provide enough environmental context for indoor navigation, and the implementation of texture mapping was non-trivial. To quickly evaluate the proposed prototype, we adopted an open 3D scanner app~\cite{3d-scanner-app} to generate textured mesh and manually aligned it to the untextured mesh with Blender~\cite{blender}.
Figure~\ref{fig:3d_map-1A} and~\ref{fig:3d_map-2A} show two 3D maps used in this study.

\subsubsection{Annotation and Format Conversion}
By zooming in, we can easily recognize large objects (e.g., trash cans) in the generated 3D maps but not small signs (e.g., room numbers). While finding a specific room is a common task in indoor navigation, we propose to use Blender to annotate the room numbers on the 3D maps to facilitate path planning. As shown in Fig.~\ref{fig:3d_map-1A} and Fig.~\ref{fig:3d_map-2A}, we add blue room numbers on the wall near the doors and the blank space behind the wall.
We also use Blender to draw distance bands on the floor on 3D maps to help agents estimate distances. As shown in Fig.~\ref{fig:3d_map-1B} and Fig.~\ref{fig:3d_map-2B}, the distance between two adjacent parallel white lines is 10 feet.

The exported 3D maps from Blender are in \texttt{.dae} (COLLADA) format. To directly load 3D maps from cloud storage in the RSA-Agent app, we converted the \texttt{.dae} file to \texttt{.scn} (SceneKit scene file) format and uploaded it to cloud storage to support online RSA services.

\begin{figure}[!tb]
\setlength{\abovecaptionskip}{-0.03cm}
\centering{
\subfigure[Example gestures to find trashcans on the first 3D map.]
{
   \label{fig:gesture-1}
   \includegraphics[width=0.98\linewidth]{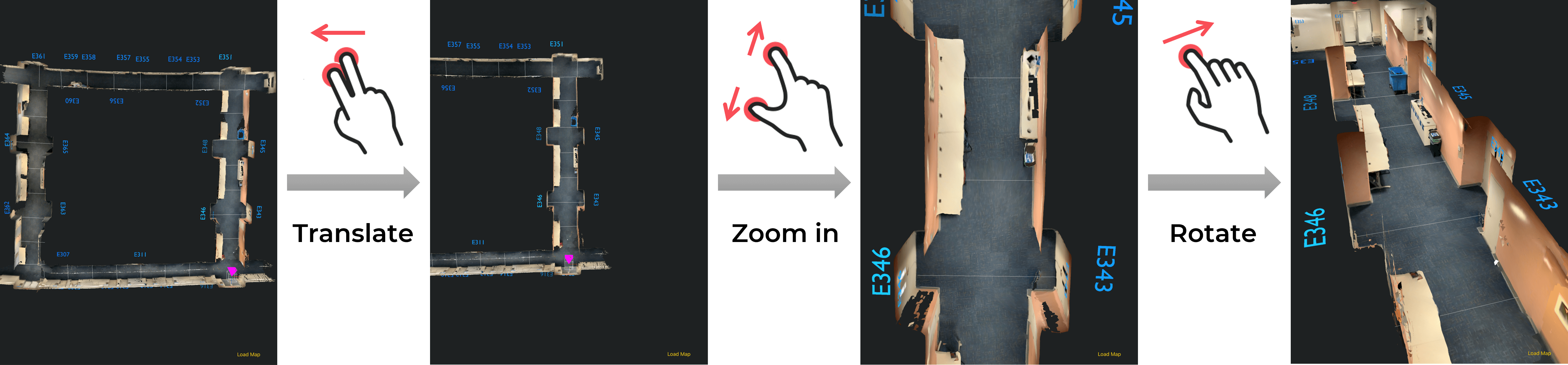}
}
\subfigure[Example gestures to find a water fountain on the second 3D map.]
{
   \label{fig:gesture-2}
   \includegraphics[width=0.98\linewidth]{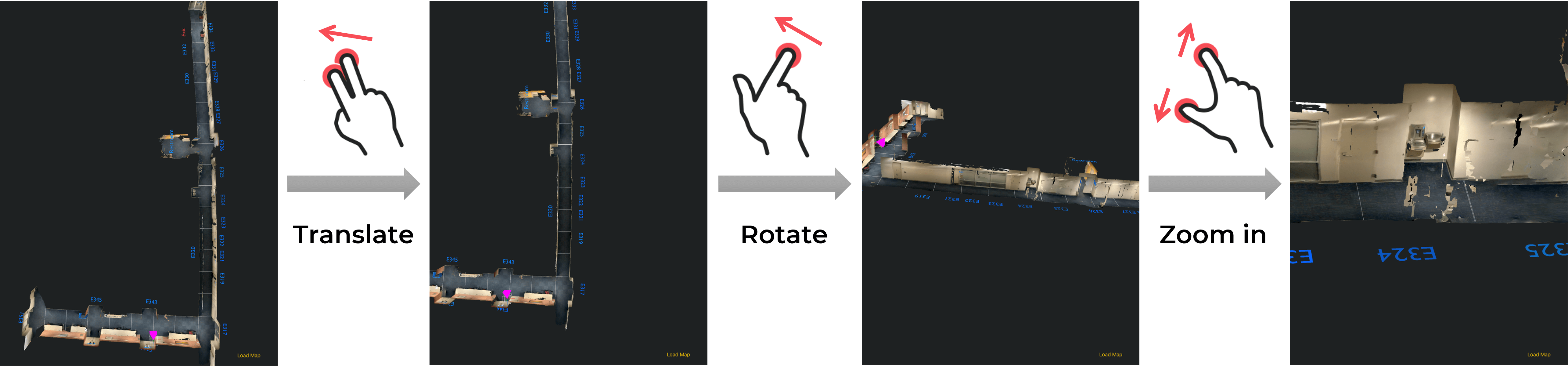}
}
}
\caption{Gesture interactions with 3D maps in our RSA-Agent app.}
\label{fig:gestures}
\end{figure}

\subsection{Online RSA}
\label{sec:online_rsa}
\subsubsection{Real-time Localization}
During RSA, the user relocalized the iPad after entering the map area. First, the user roughly scanned the environment with RSA-User app for about 10 seconds to create a new ARWorldMap object. Then, the user tapped the ``Load'' button to load the previously saved ARWorldMap file. The RSA-User app would align the two ARWorldMap objects and transform the current world coordinate system to the previous one. During the navigation, RSA-User app was continuously performing scene understanding and maintaining the high-accuracy camera pose estimated from multiple sensors including a LiDAR scanner, RGB cameras, and an inertial measurement unit (IMU).

We employ Apple's Multipeer Connectivity framework~\cite{multipeerconnectivity} to establish the connection between RSA-User and RSA-Agent apps. The user's camera pose is ceaselessly sent from RSA-User to RSA-Agent via a private Wi-Fi. In RSA-Agent app, we use a magenta pyramid-shaped camera gizmo to represent the angle of view of the user's camera, as shown in Fig.~\ref{fig:3d_map-1A} and Fig.~\ref{fig:3d_map-2A}. After loading 3D maps to RSA-Agent app, the agent can observe the camera pose (position and orientation) of the user's device on 3D maps in real-time. The RSA-Agent interface supports standard touchscreen controls provided by SceneKit to browse the 3D map: rotating the view with one-finger pan, translating the view with two-finger pan, zooming in/out with two-finger pinch or three-finger pan vertically, and resetting the view with double-tap. Figure~\ref{fig:gestures} demonstrates how to find trashcans (Fig.~\ref{fig:gesture-1}) and water fountain (Fig.~\ref{fig:gesture-2}) in the 3D maps from the top-down view with touchscreen gestures in RSA-Agent app.

\subsubsection{Live Video Chat}
The RSA-User and RSA-Agent apps currently do not support video chat. To verify the prototype, we employ off-the-shelf videoconferencing solutions to realize the live video chat. Specifically, we use Skype voice calls for user-agent oral communication. In the experiments, the user hung the iPad Pro on the chest with a belt around the neck to capture the video in front. The user's iPad screen is mirrored to the agent's laptop via the private WiFi so that the agent can see the real-time camera feed from the laptop.

\section{Experiment Design}
In this section, we will introduce the experiment design in terms of the participants, environment, apparatus, task design, and procedure. 
We aim at investigating the feasibility of interactive 3D maps for RSA, compared to the 2D maps. We assess the performance of agents with task completion time and their self-evaluation.

\begin{table}[]
\centering
\begin{tabular}{clll}
\hline
\textbf{ID} & \multicolumn{1}{c}{\textbf{Age}} & \multicolumn{1}{c}{\textbf{Gender}} & \multicolumn{1}{c}{\textbf{Occupation}} \\ \hline
P1 & 19 & F & Student (Secondary English Education) \\
P2 & 21 & F & Student (Psychology) \\
P3 & 19 & M & Student (Computer Science) \\
P4 & 18 & M & Student (Mechanical Engineering) \\
P5 & 21 & F & Student (Psychology) \\
P6 & 26 & M & Student (Mechanical Engineering) \\
P7 & 21 & M & Student (Statistics) \\
P8 & 21 & M & Student (Data Science) \\
P9 & 30 & F & Student (Information Science) \\
P10 & 20 & F & Student (Advertising) \\
P11 & 20 & F & Student (Accounting) \\
P12 & 19 & M & Student (Division of Undergraduate Studies) \\
P13 & 25 & M & Student (Industrial Engineering) \\
U* & 34 & M & Student (Information Science) \\
\hline
\end{tabular}
\caption{Participants' demographic information.}
\label{demographic_info}
\end{table}

\subsection{Participants}
We recruited 13 sighted participants on-site who were not trained for RSA previously. As shown in Table~\ref{demographic_info}, the participants were college or graduate students, ranging from 18 to 30 years old (average 21.5). There were 6 females and 7 males (self-reported gender). The backgrounds of the participants were diverse, covering natural science and social science disciplines. To verify the feasibility of the 3D maps in promoting environmental familiarity, all the participants had never been to the testing site before the experiments. Each participant was compensated for their time and effort. Our experiment is IRB-approved. 

Considering that PVI are disproportionately affected by the COVID-19 pandemic~\cite{armitage2020covid}, we did not recruit any PVI in the experiment. Instead, one sighted volunteer (marked as U* in Table~\ref{demographic_info}) acted as the user with simulated vision impairments. We blindfolded the volunteer and equipped him with a cane. Although such simulation might not perfectly show the experiences of PVI, we believe it reflects representative behavior patterns of ordinary PVI, such as being unable to get any visual information and relying heavily on verbal cues. 

Because we pay particular attention to the influence of our prototype on agents' performances, we need to minimize the confounding factors. Different users might have different O\&M skills, walking paces, and preference for more or less scenery description~\cite{c4vtochi}, which could significantly impact the agents' performances but provide less insightful design implications for our prototype. Therefore, we used the same volunteer as the user for all tasks. In addition, we also trained the volunteer to primarily rely on the instructions and make less action decision by himself.

\subsection{Environment}
The experiments were conducted in two specified office areas inside a campus building. Figure~\ref{fig:maps-1} and~\ref{fig:maps-2} show the maps of the two testing areas. 
In the experiments, each agent will be sitting in a control room that is inside the first testing area (the small room between E346 and E363 in Fig.~\ref{fig:2d_map-1}). And the user will be walking in the hallways under the agent's instructions. 
Each side of the hallways looks similar, which makes the navigation tasks (see Section~\ref{sec:task_design}) complex enough for evaluations.
The control room blocks all visual and verbal information from outside, so the agent cannot get any information about the user except from the prototype that supports online RSA. 

\subsection{Apparatus}
The user was blindfolded, equipped with a white cane and an iPad with RSA-User app installed. 
The user hung the iPad around his neck with a strap and pointed the camera of the iPad in the direction of the environment. He was able to change the orientation of the camera when needed, such as lifting it and pointing it to the left or right. 
The agents were always equipped with a laptop that enables Skype and an iPad with RSA-agent app installed. 
Depending on the type of task, we provided the participants with a 2D or 3D map on the iPad. After their consent, an iPhone was used to record agents' activities during each task.

\subsection{Task Design}
\label{sec:task_design}

Before the formal experiments, we conducted a pilot study, and it involved 9 participants who are different from the ones in the final study. 
The procedure in the pilot study is a superset of the final study.
The main outcome of the pilot study is to help us standardize the final procedures (e.g., the agent training session) and make the designed tasks manageable, reasonable, and accomplishable in one hour.

In the formal study, each agent-user pair needs to complete four tasks. Specifically, two tasks are conducted in the first condition (C1) with a 2D map, while the other two tasks in the second condition (C2) with a 3D map.
The second condition (C2) is the same as described in Section~\ref{sec:online_rsa} and Fig.~\ref{fig:workflow}, i.e., our RSA prototype with interactive 3D map, Skype for oral communication, and mirrored screen for live video.
The 3D map includes annotations of room numbers and distance bands. Thanks to the high-quality colored texture, agents are able to identify large unlabeled objects (e.g., trashcans in Fig.~\ref{fig:trashcan} and a water fountain in Fig.~\ref{fig:water_fountain}) from a 3D map by zooming in. The RSA-Agent app also shows the user's real-time location and orientation on the 3D map as seen in Fig.~\ref{fig:3d_map-1A} and Fig.~\ref{fig:3d_map-2A}.
The only difference between C1 and C2 is substituting the 3D map and RSA-Agent app with a 2D map picture on the same iPad.
As seen in Fig.~\ref{fig:2d_map-1} and Fig.~\ref{fig:2d_map-2}, the 2D maps are cropped floor plans containing basic information such as the layout of rooms, fire extinguishers, and escape routes. Both the 2D and 3D maps cover the same areas where the experiments were conducted.

We design two types of tasks for each condition: Task 1 (T1, finding a specific room with annotation on the map) and Task 2 (T2, finding an unannotated landmark).
All experiments of Task 1 were performed in the first testing area as shown in Fig.~\ref{fig:maps-1}. 
We prepared various trials for T1 with different start points and target rooms but similar distances and difficulties. The agents are free to choose any possible routes to the destination. Figure~\ref{fig:3d_map-1B} demonstrates an example of T1 starting from the green dot and ending with the green star (E351). In this example, the agent could plan the path in two different directions, either up-to-right or right-to-up.
For T2, we provide two trials of finding a trashcan (Fig.~\ref{fig:trashcan}) in the first testing area or a water fountain (Fig.~\ref{fig:water_fountain}) in the second area.
As shown in Fig.~\ref{fig:gestures}, the trashcans and water fountain can be recognized in 3D maps by changing the view with gestures. There are two groups of trashcans in the first testing area, as marked as red stars in Fig.~\ref{fig:3d_map-1B}. We regarded the task as accomplished when the user reached one of the trashcans. The start point was chosen from the two red dots in Fig.~\ref{fig:3d_map-1B}.
The second testing area was only used for T2 of finding the water fountain. Figure~\ref{fig:3d_map-2B} shows the start point (red dot) and the position of the water fountain (red star).
We assigned the two T2 trials to C1 and C2, respectively. To reduce the effect of task-specific variance, half of C1 (or C2) trials adopt finding trashcans and the other half finding water fountain.

To measure NASA-TLX separately for the two conditions, we conducted the experiments first in one condition and then in the other condition.
To minimize the influence of the order, we randomly chose half of the agents to be tested in the order C1 \& C2 and the other half in the reverse order C2 \& C1. 
For each condition, one task is from T1, while the other task is from T2.
Since there are overlaps between the testing areas of T1 and T2, we first performed T2 task then T1 task. The reason is that the agents might come across the T2 targets (e.g., trashcans) during the T1 navigation, but not vice versa; because the agents would hardly notice the target room of T1 during the T2 process.
Based on the path chosen by the agent in T2, we then specified the start point and destination of T1 in the unexplored area.
Other confounding factors were also minimized for fair comparisons of performance. For example, we only conducted the experiments at the time when there were no other people in the testing areas in case of interruptions.

\subsection{Procedure}

At the beginning of the experiment, we hosted each agent outside the testing area, introduced the study, and obtained their oral consent. We then led the agent to the control room. 
Given that the control room is located in the testing area, we covered the agent's eyes until arriving at the control room to make sure they will not get any spatial information before the experiment. We briefed the agent on the process of the experiment and taught them how to use the prototype. 
Agents were given sufficient time ($\sim$10 min) and instructions to familiarize themselves with the prototype. 
Referring to the expressions for giving guidance, we introduced agents with directional, numerical, descriptive, and status check instructions~\cite{kamikubo2020support}. Agents were encouraged to act as professionally as possible in terms of describing directional cues with contextual details and avoiding numerical terms.  
We also warned the agents about potential issues, such as latency (technical issue) and the possibility of the user walking out of the testing area (experimental issue). We emphasized that agents should be responsible for the user's safety throughout the task. To ensure the user's safety, we designated one researcher to quietly watch the experiments in the hallway.

Before each task, one researcher set up the apparatus and led the blindfolded user to a start point that was unknown to both the agent and the user. After arrival, the researcher told the user about the specific task (e.g., finding an office or a water fountain). 
We defined that a task started when the agent and the user began communication and ended when the user successfully found the target location or object. With their consent, we filmed the agents' activities during each task. We also documented the completion time of each task and double-checked from the recorded videos.

We aim at investigating and comparing the agents' experiences in two different conditions. Therefore, each agent was asked to perform two tasks in one condition first. After finishing the tasks, we measured the agent's perceived workload using the NASA-TLX. 
Next, the agent conducted the other two tasks in the other condition and assessed their workload with NASA-TLX. The NASA-TLX scores were recorded and used as a prompt in the following interview. Finally, we conducted a semi-structured interview with the agent. The interview questions were mainly about the experiences of guiding the user and comments on the two conditions. The interviews were audio-recorded and transcribed into text documents. Each session lasted for about 90 minutes.

\section{Quantitative Analysis}

\begin{figure}[!tb]
\setlength{\abovecaptionskip}{-0.03cm}
\centering{
\subfigure[Task completion time]
{
   \label{fig:completion_time}
   \includegraphics[width=0.39\linewidth]{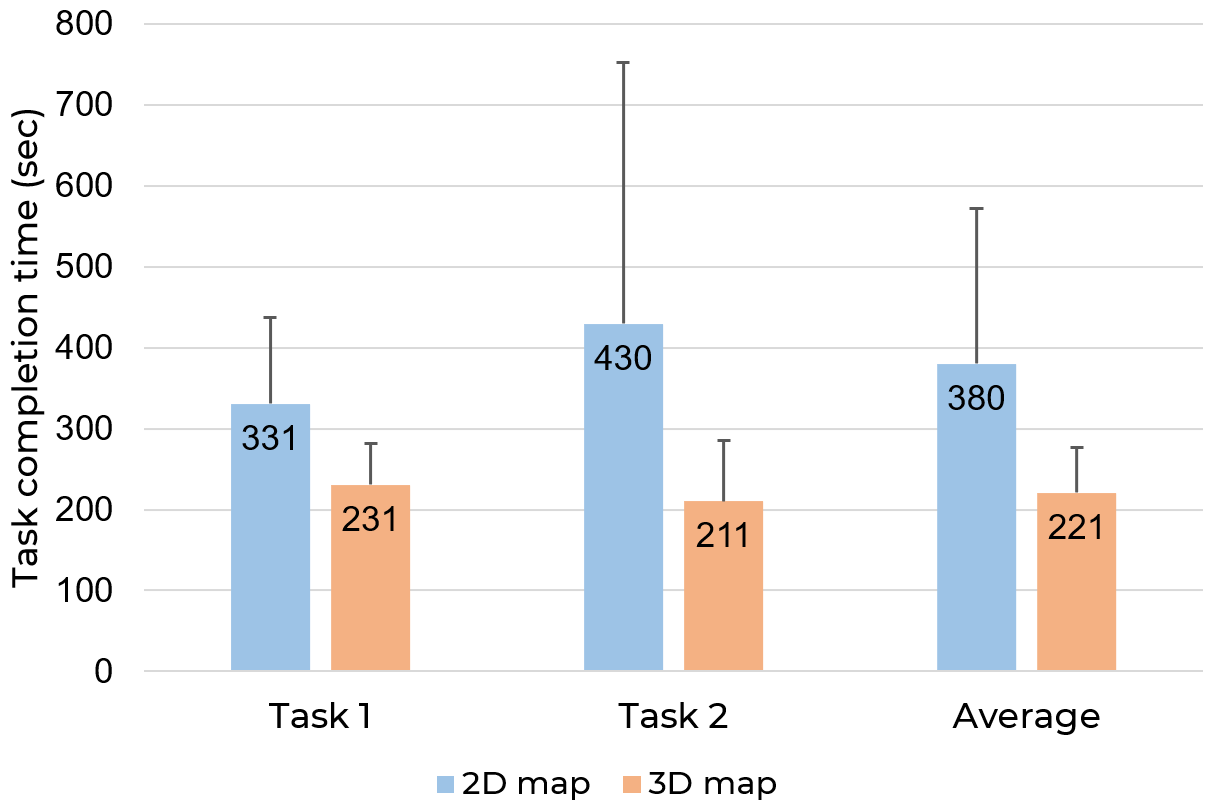}
}
\subfigure[NASA-TLX scores]
{
   \label{fig:nasa-tlx}
   \includegraphics[width=0.59\linewidth]{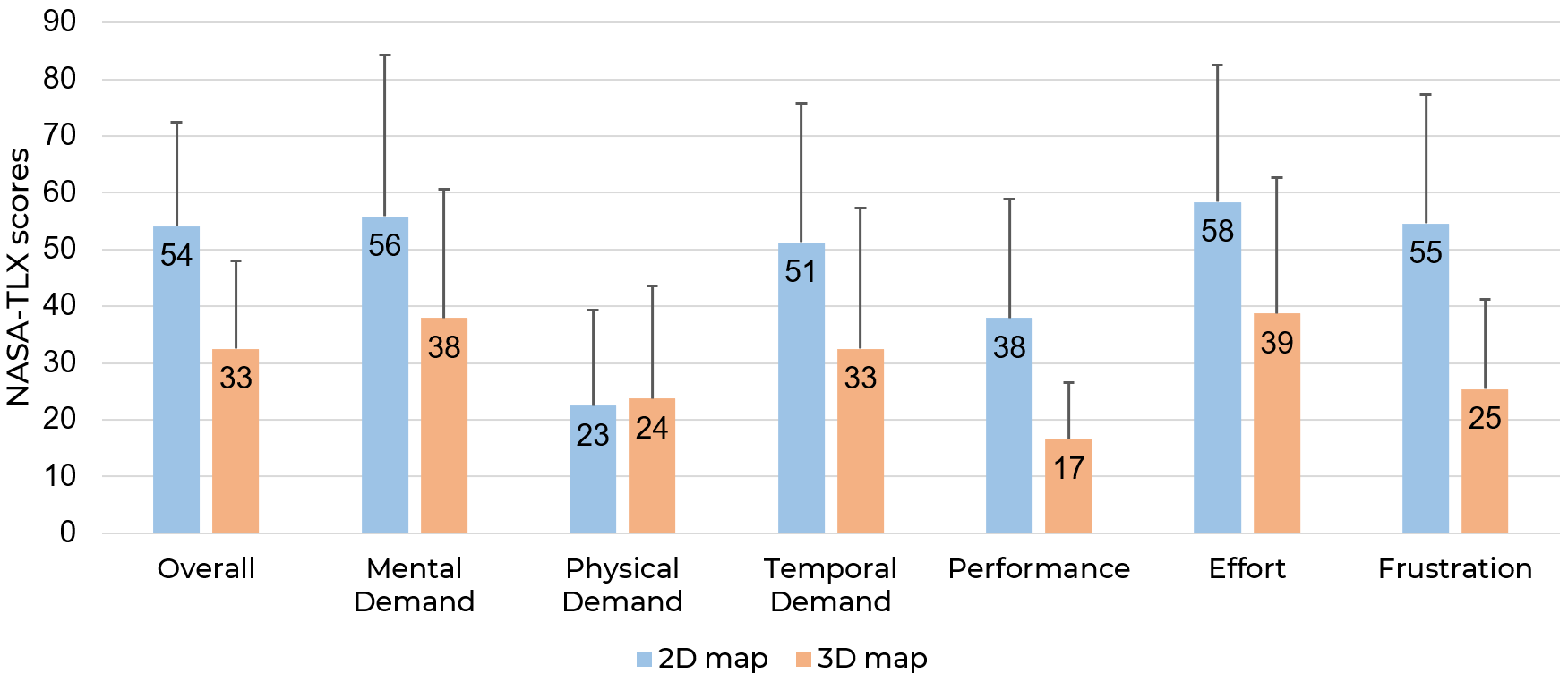}
}
}
\caption{Comparisons between 2D and 3D map in (a) task completion time and (b) NASA-TLX scores.}
\label{fig:quantitative}
\end{figure}

In this section, we will analyze two sets of data quantitatively, including (i) task completion time and (ii) subjective evaluations with NASA-TLX scores. 
One outlier was detected in the dataset of task completion time by both $z$-scores and interquartile ranges. P13 spent a much longer time in finding an unannotated landmark with 3D maps than other participants. The reason was that he didn't realize the real-time localization and orientation in the first trial of using 3D maps \textit{``so I [he] just did the task in the way I [he] did with the 2D map''}. After realizing the functionality of 3D maps, he reported that \textit{``the last task I followed the red spot so I think I finish the task very quickly''}. 
Thus, we removed P13's task completion time. 
Considering that P13's unawareness of 3D map functionality could affect his self-evaluation, we removed his NASA-TLX scores as well. A paired, two-tailed $t$-test was computed on the remaining data sets.

\subsection{Task Completion Time}
\label{completion_time}
Figure~\ref{fig:completion_time} shows the mean completion time for (i) Task 1, finding a room with annotation on the map, (ii) Task 2, finding an unannotated landmark, and (iii) both tasks. The completion time of using 3D map was significantly shorter than that of using 2D maps in Task 1 ($p = 0.001$), Task 2 ($p = 0.034$), and average of both tasks ($p = 0.008$). The results indicate that 3D maps can significantly reduce the task completion time in finding either annotated or unannotated targets. 
Especially in Task 2 of finding unannotated landmarks, the mean completion time of using a 2D map ($\overline{t}=430 \sec$) is more than twice that of using a 3D map ($\overline{t}=211 \sec$). It demonstrates the superiority of 3D maps in finding landmarks that are not annotated but can be recognized from the 3D structure and texture.

\subsection{Subjective Results} 
\label{tlx_results}
NASA-TLX scores elicit agents' experiences in terms of (i) overall experiences, (ii) mental demand, (iii) physical demand, (iv) temporal demand, (v) performance, (vi) effort, and (vii) frustration. Figure~\ref{fig:nasa-tlx} shows that the mean NASA TLX scores of using 3D maps were lower than that of using 2D maps in all measures, except for physical demand. In particular, the mean performance and frustration scores of 3D maps are only around 45\% of that of 2D maps, indicating 3D maps can considerably improve the self-rated performance and reduce the frustration of agents in indoor navigation tasks. It is also understandable that 3D maps received slightly higher scores in physical demand ($\overline{x}=24$) than 2D maps ($\overline{x}=23$), because interactive 3D maps require extra gesture interactions compared with static 2D maps.

The results of $t$-test are consistent with Fig.~\ref{fig:nasa-tlx}. 
3D maps help the agents significantly reduce the mental demand ($p = 0.013$), temporal demand ($p = 0.024$), effort ($p = 0.011$), and frustration ($p < 0.001$). The agents also believed they performed significantly better with 3D maps than 2D version ($p = 0.002$). 
Moreover, there are no statistically significant differences between 2D and 3D maps in agents' physical demand ($p = 0.699$), suggesting that interactive 3D maps did not introduce additional physical burdens compared with 2D maps.

In sum, we found from the experiments that 3D maps could significantly improve the agents' performance in task completion time and by the subjective assessment, and reduce their mental demand, temporal demand, effort, and frustration while not markedly increasing the physical demand.

\section{Qualitative Analysis}
We conducted semi-structured interviews with the agents after they finished all the tasks. We elicited their experiences of instructing people's indoor navigation. We used thematic analysis \cite{Braun2006b} to understand such experiences. First, the authors went over all transcripts to identify data related to the agents' experiences in both conditions. For example, how did the agents used a 2D/3D map to identify where the user was. The authors open-coded such experiences without any pre-defined framework. Second, the authors held regular meetings to refine and organize the codes. The meetings focused on discovering the underlying relationship between codes; codes that were closely related would be grouped together. For instance, we identified that some codes were about agents' experiences of planning the path for the user; we then grouped such code together and named the group as ``navigational strategy''. We identified themes through the process; we also went back and forth to testify if the themes were mutually exclusive and applicable to most codes. Third, we came up with three themes and finalized their names as ``navigational strategy'', ``efficient coordination with reducing unnecessary instructions'', and ``user experience of split-screen dashboard''. We reported the themes in detail in the following subsections.

\subsection{Navigational Strategy}

During the tasks, agents utilized navigational strategy to make sense of the spatial layout, path planning, and the user's location and orientation. Common examples of navigation strategies are leveraging real-time updates and detailed visualization of the environment.  

\subsubsection{Real-time Localization and Update}
\label{real-time_update}

During the experiments, real-time data was always captured and provided from the live camera feed. The other type of real-time data was represented as real-time localization and orientation, which was only available on the dynamic and interactive 3D maps but not on the static 2D maps.

Getting aware of the starting point, that is, where the user is at the beginning of the tasks, is the first step and challenge in navigation. The success and easiness of this step affect agents' subsequent decision-making and confidence. 
Seven agents (P1, P3, P5, P8, P9, P11, P13) indicated that real-time localization on 3D maps helped them make sense of the user's location throughout the navigation. Compared with 2D maps, it \textit{``is the most helpful thing''} (P5) and \textit{``makes everything easier''} (P8). P10 thought that the real-time localization was an extra help that got rid of the reliance on scanning and broke the restriction of limited camera view.
\begin{quote}
    \textit{``Because you can see the actual thing, both on the [3D] map and his camera. So like I said, I can combine these things together to imagine where he actually was, but the 2D map, I can only rely on his camera, which is really limited.'' \textup{(P10)}}
\end{quote}

Orientation is another important factor in navigation. Even if agents localize the user correctly, failure to identify the orientation can result in navigating the user in the wrong direction and leading the user away from the destination. Deducing the user's orientation is more challenging in homogeneous environments. As pointed out by P10, \textit{``the roads, the walls, the hallways are kind of similar''} so she made a wrong judgment of direction without 3D maps. 
Identifying the user's orientation requires mental efforts because P8 \textit{``need[ed] to put myself [himself] in a location and to check if that's the right direction''}. Additionally, he had to \textit{``keep thinking every time he [the user] made a turn''}. 
Real-time update of the user's orientation on 3D maps alleviates this problem. Benefiting from this feature, P1's cognitive load was reduced by not putting efforts to situate herself in the user's physical position and \textit{``translate''} the user's orientation.

\subsubsection{Detailed Visualization of the Environment}
Visualization of the environment is represented as the room numbers, fire extinguisher signs, and building layout on 2D maps, as well as the room numbers, distance bands, and detailed environmental information on 3D maps. Compared with the abstract spatial layout on 2D maps, 3D maps supplement more environmental details, such as the location, shape, and color of objects. These details are not limited to annotated targets as on the 2D maps, but all the objects in the scope of the testing area. This feature of 3D maps increased agents' familiarity with the environment because they \textit{``can see everything on the map''}. 
\begin{quote}
    \textit{``Because for the 3D map, you can see everything... like tables, desks, water fountain, and trash cans, you can see everything on the map. But for 2D map, not that many details.'' \textup{(P7)}}
\end{quote}

This characteristic makes 3D maps not only more comprehensive but also more intuitive. As pointed out by P9, reading 2D maps is challenging because she needs to transform the spatial layout in mind and match it to 3D scenes that she is familiar with. The detailed environmental information on 3D maps is \textit{``natural''} as it is similar to what people absorb and experience in their daily lives. 
\begin{quote}
\textit{``I'm not good at reading [2D] map. So reading map itself is hard for me... I'll get lost in the map because I'm really scared of reading [2D] map... [But] the objects are all on the 3D map... I think it's just quite natural to use the tool [3D map].'' \textup{(P9)}}
\end{quote}

As mentioned in Section~\ref{real-time_update}, agents aligned the real-time data from the live camera feed with the information on 2D maps for localization and orientation. In addition to these purposes, visualization of the environment on either 2D or 3D maps facilitates agents to find the destination, which is crucial in path planning and agents' decision-making. 
Ten agents (P2, P3, P5, P6, P7, P8, P9, P10, P11, P12) believed that the detailed environmental information on 3D maps made the wayfinding easier, especially in the task of finding an unannotated object (e.g., trash can or water fountain). The reason was that they could be more familiar with the environment by identifying the object with its shape or color and get aware of its location on 3D maps. 
The determination of the destination accelerated the navigation process by contributing to accurate decision-making and reducing the possibility of giving wrong instructions. It decreased agents' mental demand and made them feel more confident and \textit{``secure''} (P7).  
\begin{quote}
    \textit{``The water fountain is actually on the 3D map. So I just look around the map and then just choose the one I want him to go to. And then just ask him to go there... Because you can scroll the [3D] map, you can just see where the direction really is. And you can just move the [3D] map to see the surrounding environment. So that may help me to make more accurate decisions.'' \textup{(P10)}}
\end{quote}

In comparison, the lack of detailed environmental information on the 2D maps led to frustration when finding an unannotated object. 
Agents had limited environmental knowledge and familiarity in this task because they were unable to find the specific object on 2D maps and further determine the destination. The only strategy was to guide the user to randomly explore the testing area.  
\begin{quote}
    \textit{``...he says he wants to use the water fountain but [it’s] not shown on the map. So the only way I know where it is is just to get him to walk around until I find that. So that is like a big frustration. You don't know where the destination is.'' \textup{(P3)}}
\end{quote}

\subsection{Efficient Coordination with Reducing Unnecessary Instructions}
In addition to making strategies for navigation, agents further reported their experiences of coordination. Because the goal of the tasks was to help the user find a location or an object, agents needed to digest the information perceived through navigation strategies and convert it into feasible instructions for the user. Therefore, the navigation tasks can be considered as a type of coordination between the agent and the user. Especially for the agents, they needed to make sense of the environment and give clear, accurate, and timely feedback to the user. Such challenges made agents use the best of the split-screen dashboard to facilitate coordination. 
Overall, agents found the 3D map was more helpful than the 2D map while giving instructions to coordinate with the user.

Localize and orientate the user was an indispensable part of giving instructions. With the 2D map, the only source of this information was the camera feed. In this condition, localization and orientation involved tedious work by asking the user to move back and forth and to adjust the angle of the camera, so that the agent could get a clearer view from the camera and get more accurate information. P8 complained that such work was effort-consuming. In contrast, 3D maps showed the location and orientation of the user, significantly reducing the agent's effort and time of instructing PVI to scan the appropriate areas at the correct angle. 

\begin{quote}
    \textit{``[When using 2D map] Because sometimes he's not facing directly to the number of the door. So that requires lots of work. I asked him to turn right, turn left, and then maybe slightly right, turn up, go back, check number. Yeah, that this process is adding more effort for me to know the location. But 3D map shows where he's facing to... I didn't really ask him to turn to check the doors or something like that. So the 3D map saves lots of time by showing the location.'' \textup{(P8)}}
\end{quote}

In addition to the localization and orientation, the environmental information on 3D maps provided a better condition for exploration. In some trials, agents were asked to find unannotated objects, such as a water fountain. Thus, agents had to explore the space to look for the objects. The exploration process required the agents to coordinate with the user with the help of the dashboard. The map became an important source of information for the exploration, and the information provided by the map determined the efficiency of the coordination. 
P5 recalled her experiences of looking for something that was not annotated on both maps:   
\begin{quote}
    \textit{``On the 2D map, I cannot find it [water fountain] because I don't know where the sign is. I instruct him to go the wrong way so I feel frustrated at that time. But on the 3D map, it is clear to see all the things on the map, so I think I don't make any wrong instructions. So I feel good.'' \textup{(P5)}}
\end{quote}

The absence of environmental information on 2D maps negatively impacted P5's navigation because she guided the user in the wrong direction and felt bad about it. However, she noticed that, although the objects were not annotated on the 3D map as well, she could identify the water fountain from the detailed visualization of the environment displayed on the 3D map. The detailed environmental information on the 3D map made up for the shortcomings that not all objects were annotated. It also reduced the frustration of P5 and made her feel more confident when giving instructions.

Agents also illustrated the benefit of the 3D map in terms of monitoring the user's movement. P10 reported why she paid constant attention to the user's movement and how she did that with both the 2D and 3D maps:

\begin{quote}
    \textit{``The 3D map helped me to define his step and the distance between him and those other things. But when I was using a 2D map to guide him to a room, I just asked him to turn left at a time when I thought he should turn left, but actually he was not at the actual place. Because when he turned left, I saw he would hit the wall. So that is a problem. And I think the 3D map helped me to better define where he was.'' \textup{(P10)}}
\end{quote}

The testing area consisted of many intersections, where the user needed to make a turn to get to the destination. To agents, deciding where and when to make turns was another important part of instructing the user. Therefore, agents should precisely measure how fast the user walked and how long the user needed to walk until the next turn. Failing to do so might lead the user to the wrong place and sometimes even expose him to risks. Due to the lack of a real-time update on the 2D map, agents were unable to get the accurate location and orientation of the user. P10 pointed out that it was difficult to measure the pace of the user through the camera. Thus, she complimented the real-time localization and environmental information on 3D maps. These features helped P10 better predict when the user would reach the intersection and then give timely directions to the user.

Lastly, the coordination involved correcting wrong directions. Our experiment took place in a part of a building, so it was likely that the agent would give wrong directions to the user and make him walk outside of the testing area. Therefore, agents were also responsible for identifying whether the user was out of the area and guiding him back if he was. P5 told us when she realized the user might have walked out of the testing area:    

\begin{quote}
    \textit{``Because I need to ask him to use the camera to see the room number [when using 2D map]. But when I use the 3D [map], I can see from the map instead of asking him...I can know he goes the wrong way. So I can make a quick decision to ask him to go around or something like that.'' \textup{(P5)}}
\end{quote}

Before the experiment, we trained each agent, telling them that the area was small and the user might walk out. The 3D map's localization function helped cope with such situations, said P5. It could instantly show that the user was going in the wrong direction. Then, the agent was able to give immediate instructions to let the user turn back to prevent him from walking out of the testing area. But when using 2D maps, P5 could not get such timely feedback; she had to ask the user to turn around and get information from the camera. If the user was near a room that was not on the maps, then he was out of the area. Such comparison indicated that the 3D map was better in terms of correcting and preventing mistakes.

\subsection{User Experience of Split-screen Dashboard}

Agents were split on their experience of interacting with the split-screen dashboard. 
Some agents appreciated the interactivity of 3D maps, which made RSA \textit{``more playable and less frustrated''} (P8). 
Other agents reported the usability issues of 3D maps, such as unfamiliarity with complicated manipulation and latency.

\subsubsection{Interactivity}
The interactivity of 3D maps is one of the characteristics that distinguish 3D maps from 2D maps. P8 said that the interactivity of 3D maps made the RSA interaction \textit{``more playable and less frustrated''} compared with 2D maps. He considered the navigation with 3D maps was \textit{``a fun thing to do''} rather than a job because he can \textit{``play with a map''}. More specifically, P2 thought the interactivity of 3D maps was represented as cuteness, where the sign of real-time localization and orientation was jumping and \textit{``moving like a duck''}. This characteristic of 3D maps released her frustration and made her more willing to help the user. The reduction in frustration was also reflected in Section~\ref{tlx_results}. 
\begin{quote}
    \textit{``I think the 3D map is cute. Because I saw that red point moving like a duck. When I directed that person moving, I thought that point was like jumping... When I thought the point, I just couldn’t help laughing. It just releases my frustration and I’m more willing to take time to help that person to get to the right place.'' \textup{(P2)}}
\end{quote}

\subsubsection{Usability}
Some agents were bothered by the complicated manipulation and connection issues when using 3D maps. Regarding the manipulation, P12 said that \textit{``it's a little bit hard for the user who doesn't get used to those kinds of three-finger thing''} and \textit{``to memorize all the steps''}. Some agents (P1, P2, P7) spent time learning and familiarizing themselves with the manipulation. P2 thought that the manipulation was hard at first, but after practice, she adjusted to it and found it easier.

Due to the unfamiliarity with the manipulation, two agents (P11, P13) preferred 2D maps rather than 3D maps. They indicated that 2D maps were more straightforward and required less manipulation. 
Although P11 acknowledged that the real-time update on the 3D map helped her localize the user, she needed to rotate the map to a certain angle to find the annotations. In contrast, annotations on 2D maps are obvious and easy to obtain. 
\begin{quote}
    \textit{``I feel like the 2D map is easier than the 3D map, because it's easier to locate the room number. I'm not too familiar with the 3D map. The 3D map is still clear, but if you want me to choose, I would prefer the 2D map over the 3D map.'' \textup{(P11)}}
\end{quote}

Latency is the other issue that affects the usability of the prototype. Five agents (P3, P5, P6, P10, P12) have encountered the connection issue during the tasks. P12 spent time and mental efforts in identifying the user's current location under this circumstance. Similarly, P6 indicated that the latency on 3D maps was \textit{``misleading''}, which confused him about the user's location and made him nervous about delivering wrong instructions. 
\begin{quote}
    \textit{``However, the problem is there's some latency on the map. I only notice once, it wasn't very representative where the person was on the map because it moved rapidly fast from one location to the next... It was a little bit misleading to me... I want to get rid of wrong commands. That's why I was nervous.'' \textup{(P6)}}
\end{quote}

Three agents (P3, P5, P10) utilized the cues from the live camera feed to mitigate this challenge. They indicated that the real-time update from the live camera feed was accurate and reliable for localization and orientation, which could compensate for the latency on 3D maps.

\section{Discussion}

In this section, we identified several advantages and opportunities of interactive 3D maps. We discuss the functionalities of 3D maps that contribute to the decreased task completion time and the design insight based on interactivity and manipulation. Finally, we present the limitations of our research and the directions for future work.

\subsection{Better Performance with Completion Time}

Our quantitative analysis shows that interactive 3D maps can significantly shorten the task completion time. Findings from the qualitative analysis are consistent with the quantitative results and supplement more details about how the functionalities of the 3D map contribute to the reduction of the time of guidance. 

First, real-time localization capability undertakes the task of localizing the user by updating the user's location throughout the navigation. It directly reduces the time and effort of determining the start point, which is the first step and challenge in navigation. 
Second, the synchronized orientation on the 3D map decreases the probability of orientating and navigating the user in the wrong direction, especially in homogeneous environments. 
Third, detailed visualization of the real world increases agents' environmental familiarity as they \textit{``can see everything on the map''}, including the location, shape, and color of both annotated and unannotated objects. In the task of finding unannotated objects, detailed visualization information also facilitates the determination of the destination and further accelerates the navigation process. 
When assisting the user with 2D maps, agents need to guide the user to scan the appropriate areas at the correct angle to find semantic information or visual cues of interest. Obtaining real-time data through the live camera feed and align it with the information on 2D maps is the only strategy to localize and orientate the user, which makes the task quite challenging and time-consuming. 
Compared with 2D maps, the aforementioned capabilities of 3D maps reduce the requirement for precise scanning of the user's surroundings and further get rid of the lengthy back and forth conversation.

\subsection{Interactivity and Manipulation}

Some agents have positive experiences in using 3D maps. They appreciated the interactivity of 3D maps and indicated that this characteristic made the navigation \textit{``more playable and less frustrated''} (P8). One representative of interactivity is cuteness. It kept P2 engaged in the RSA interaction because she became \textit{``more willing to take time to help that person to get to the right place''}. 

Compared with 2D maps, the interactivity of 3D maps involves manipulating the maps with gestures. Some agents considered the manipulation as complicated and were bothered by extra attention and mental demand to memorize the gestures and familiarize themselves with them. Two agents (P11, P13) even preferred 2D maps because they are more straightforward and require less manipulation. This informs the design insight for the next version of our prototype. We can improve the interactivity with different input modalities. For example, we can use the ``left'' or ``right'' button to replace with the two-finger pan for translating the view, use the ``+'' or ``-'' button to replace the two-finger pinch for zooming in/out. Additionally, we can add shortcut buttons to switch between different views (e.g., first-person view and top-down view) instead of manually rotating the map.

\subsection{Limitations and Future Work}
\paragraph{\textbf{\textit{Participants}}}

Our findings show that the interactive 3D maps alleviate the navigational challenges for untrained agents by supplementing the environmental knowledge, real-time localization, and orientation. Trained agents are also bothered by these challenges~\cite{kamikubo2020support}. Thus, we believe that interactive 3D maps have the potential to be a boost for trained RSA agents. We acknowledge the limitation of no trained agents involved in the study. Our future direction is to review study footage with Aira agents for expert evaluation.

We understand that RSA is a collaborative interaction between PVI and agents. The RSA users with visual impairments receive assistance from agents, but at the same time, they help agents with their knowledge about the area/landmarks and expertise in O\&M skills using a cane or a guide dog~\cite{lee2020emerging}. This expertise of PVI in blind navigation could affect the performance of our untrained agents. It is a limitation that no PVI participate in this study. 
The prototype is still in the research phase and it has not been tested to be fault-tolerant. We will engage PVI in the future while ensuring their safety and lowering the risk.

\paragraph{\textit{\textbf{Prototype}}}
It is worth noting that our RSA system is only a prototype for assessing the feasibility of 3D maps for indoor navigation. There are several features that need to be further improved in our future implementation. First, the video chat is currently realized by Skype and screen mirror. We will integrate the video chat function to RSA-User and RSA-Agents apps. Second, the initialization of RSA-User app (e.g., loading 3D map) was conducted by a sighted person for the user in this study. For practical use, we need to make RSA-User app more accessible, e.g., with voice input for initialization. Third, we will implement texture mapping in our RSA-User app to generate high-quality textures for 3D maps.

Another limitation of the current implementation is latency and localization error. 
The issues of latency may arise due to unstable connections. Though ARKit-based localization is accurate enough in the experiments, one of the main adverse effects of latency was to display the real-time location of the user with lags.
In our study, five participants (P3, P5, P6, P10, P12) encountered the latency issue during the experiments, but three of them (P3, P5, P10) overcame this challenge by associating 3D map with live camera feed.

Our RSA prototype is also not ready to be deployed in large-scale scenarios. First, the current implementation employs Apple's Multipeer Connectivity framework for data transmission between RSA-User and RSA-Agent apps. The framework only works in a WiFi network and doesn't support out-of-range remote communication. Second, we only tested the apps with 3D maps of moderate size. When applied to large buildings (e.g., grand shopping malls), the current prototype may experience problems in processing the 3D map. We expect these issues will be addressed in the future with the upgraded mobile devices and the advent of the 5G era.

\section{Conclusion}

We designed a prototype to assess the feasibility of interactive 3D maps in addressing the indoor navigation challenges in RSA. In line with existing RSA services (e.g., Aira), our prototype is developed for PVI’s mobile devices without extra setups. Thanks to ARKit and LiDAR scanner, we are able to create high-quality 3D maps and provide accurate localization with an iPhone or iPad. In walkthrough evaluations, the results show that 3D maps can significantly improve RSA agents’ performance in indoor navigation tasks. The features of real-time localization, landmark annotations, and fine-grained 3D map details are favored by most of the participants. 
In this study, the prototype implementation and empirical results provided concrete evidence supporting the feasibility of 3D maps for RSA.
We hope that our study can pave the way to enhance RSA systems with 3D maps for indoor navigation.

\bibliographystyle{style.bst}  
\bibliography{references}

\end{document}